\documentclass[english]{iopart}

\usepackage{iopams}
\usepackage{babel}
\usepackage{units}
\usepackage{amsbsy}
\usepackage{amssymb}
\usepackage{graphicx}
\usepackage{esint}
\usepackage{hyperref}

\makeatother

\usepackage{subfigure}  
\begin{document}

\newcommand{\gtwid}{\mathrel{\raise.3ex\hbox{$>$\kern-.75em\lower1ex\hbox{$\sim$}}}}
\newcommand{\ltwid}{\mathrel{\raise.3ex\hbox{$<$\kern-.75em\lower1ex\hbox{$\sim$}}}}

\title[]{Quasi-Particle Interference Probe of the Self-Energy}

\author{Thomas Dahm}
\address{Universit\"at Bielefeld, Fakult\"at f\"ur Physik, 
Postfach 100131, D-33501 Bielefeld, Germany}

\ead{thomas.dahm@uni-bielefeld.de}

\author{D.J.~Scalapino}
\address{University of California, Physics Department, Santa Barbara, CA 93106-9530, USA}

\ead{djs@physics.ucsb.edu}


\begin{abstract}
Quasi-particle interference (QPI) measurements have provided a powerful tool for
determining the momentum dependence of the gap  of unconventional
superconductors. Here we examine
the possibility of using such measurements to probe the frequency and momentum
dependence
of the electron self-energy. For illustration, we calculate the QPI response function for
a cuprate-like Fermi surface with an electron self-energy from an RPA
approximation.
Then we try to reextract the self-energy from the dispersion of the peaks in
the QPI response function using different approaches. We show that in principle
it is possible to extract the self-energy from the QPI response for
certain nested momentum directions. We discuss some of the
limitations that one faces.
\end{abstract}

\pacs{74.55.+v, 74.25.Jb }

\maketitle

\section{Introduction}\label{sec:1}

Useful information about the interaction of electrons in metals and superconductors
is contained in the quasiparticle self-energy $\Sigma(k,\omega)$.
When the self-energy and the band structure $E(k)$ of a metal is known, the
Green's function $G(k,\omega)=\left[ \omega - E(k) - \Sigma(k,\omega) \right]^{-1}$
provides complete information on the single particle properties of the system.
Experimentally, in isotropic superconducting systems both the normal and
anomalous (gap function) self-energies at the Fermi surface can be obtained
from tunneling spectroscopy. \cite{McMillan}
In anisotropic systems angular resolved photoemission spectroscopy (ARPES)
can be used to obtain momentum resolved information on the
self-energy. \cite{Valla,Bogdanov,Kaminski,ARPESMDC,Yun,Zhang}
In recent years it was shown that the momentum dependence of the gap
in unconventional superconductors can be obtained from scanning tunneling
microscopy (STM) employing the so-called quasi-particle interference (QPI).
\cite{Byers,ref:1,ref:2,ref:3,Hanaguri2,ref:4,ref:5,Fujita,Allan2}
In this technique one measures the local tunneling conductance at $\omega=eV$ around an
impurity at the surface of a metal over a large grid of
points. Its Fourier transform gives a wave vector power spectrum
$|{\rm Im}\,\Lambda(q,\omega)|^2$. Peaks in ${\rm Im}\,\Lambda(q,\omega)$
arise from dynamic nesting processes in which quasi-particles undergo elastic
backward scattering from the impuities. From these peaks and their dispersion
one can obtain information on nesting properties of the Fermi surface and
about the momentum dependence of the superconducting gap. \cite{Capriotti}
In the present work we want to explore, whether beyond that QPI experiments
can be used to also extract information about the momentum and energy dependence
of the self-energy, in particular about the effective mass and the lifetime
of the quasiparticles. The main idea here is to closely investigate the
dispersion and the width of the peaks in ${\rm Im}\,\Lambda(q,\omega)$
as a function of energy $\omega$ and try to extract the self-energy from it.
We will demonstrate that in principle this is possible and we will discuss some of
the limitations that one faces.

In a metal, the dispersion and damping of quasi-particles with energy $\omega$
is described by a complex frequency dependent wave vector $k(\omega)=
k_1(\omega)+ik_2(\omega)$, where $k_1(\omega)$ determines the renormalized dispersion
and $k_2(\omega)$ the lifetime of a quasiparticle state.
 The band structure $E(k)$ and the self-energy
$\Sigma(k,\omega)$ determine $k(\omega)$ and conversely, given $E(k)$, the
structure of the self-energy is reflected in $k_1(\omega)$ and
$k_2(\omega)$. The tunneling conductance at a particular point depends
upon $k(\omega)$ and the surrounding impurity configuration. 
Here we will discuss how one can extract
$k(\omega)$ from the structure in ${\rm Im}\,\Lambda(q,\omega)$
and use it to study the real and imaginary parts of the self-energy.

\section{Peaks in the QPI response}

\begin{figure}[htbp]
\begin{center}
\includegraphics[width=0.45 \columnwidth]{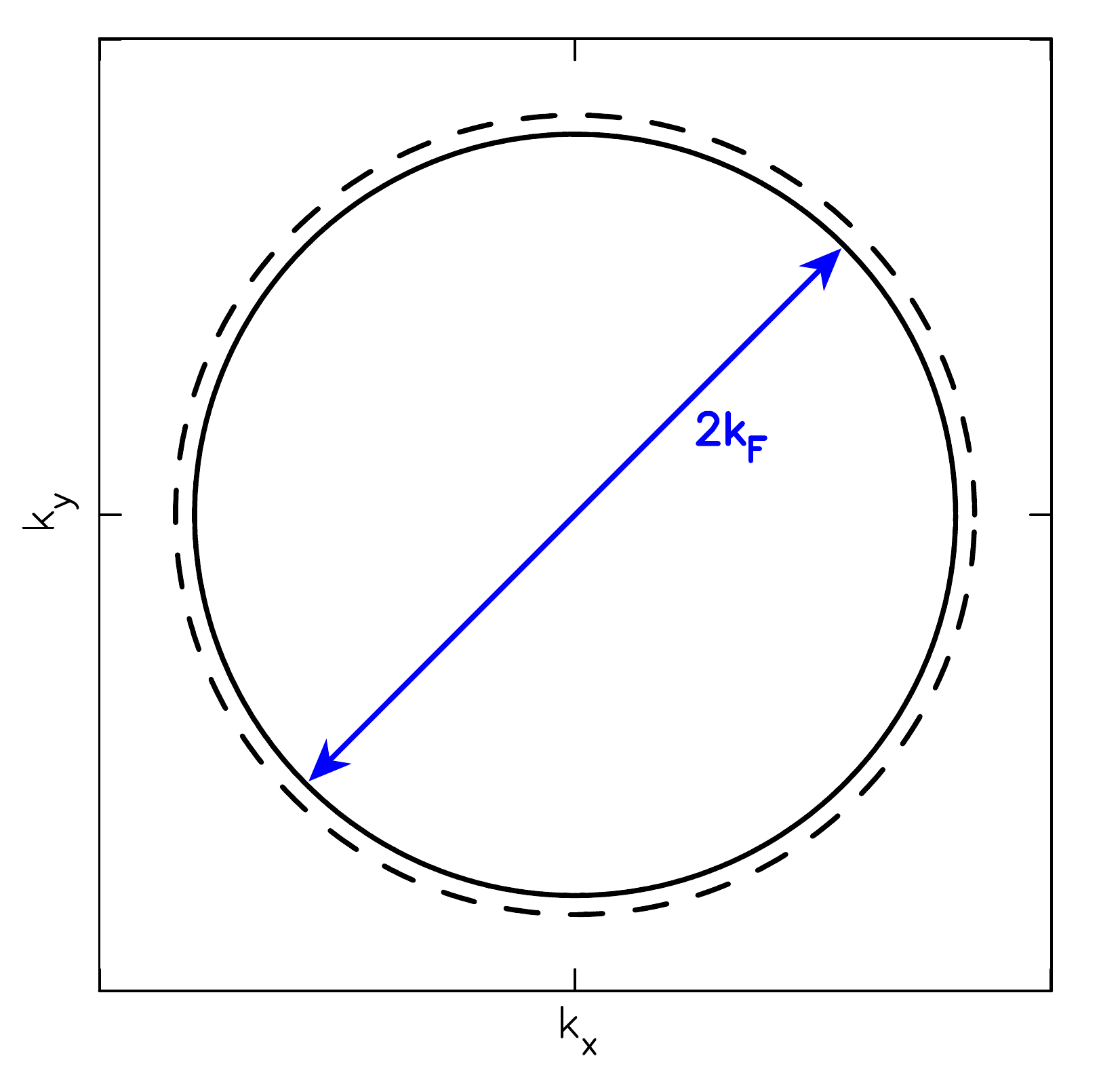}
\end{center}
\caption{A $q$-cut passing through the Fermi surface of a 2D free electron
system. For $\omega=0$, $q=2k_F$ leads to nesting, while for $\omega=0.1\mu$
(dashed circle) nesting occurs for $q=2k(\omega)=2k_F\sqrt{1.1}$.\label{fig:1}}
\end{figure}

\begin{figure}[htbp]
\begin{center}
\subfigure{
\includegraphics[width=0.45 \columnwidth]{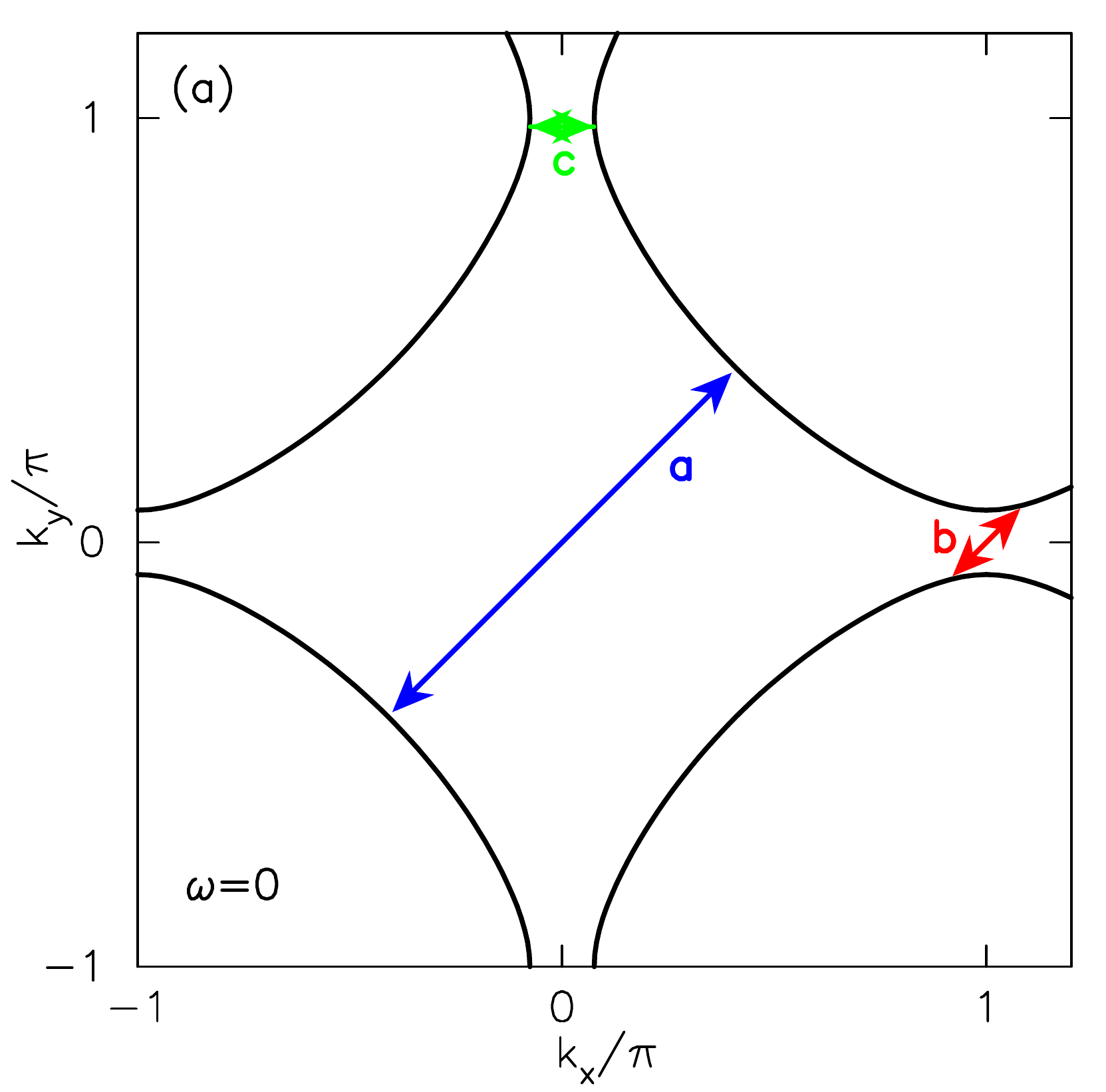}} \\
\subfigure{
\includegraphics[width=0.47 \columnwidth]{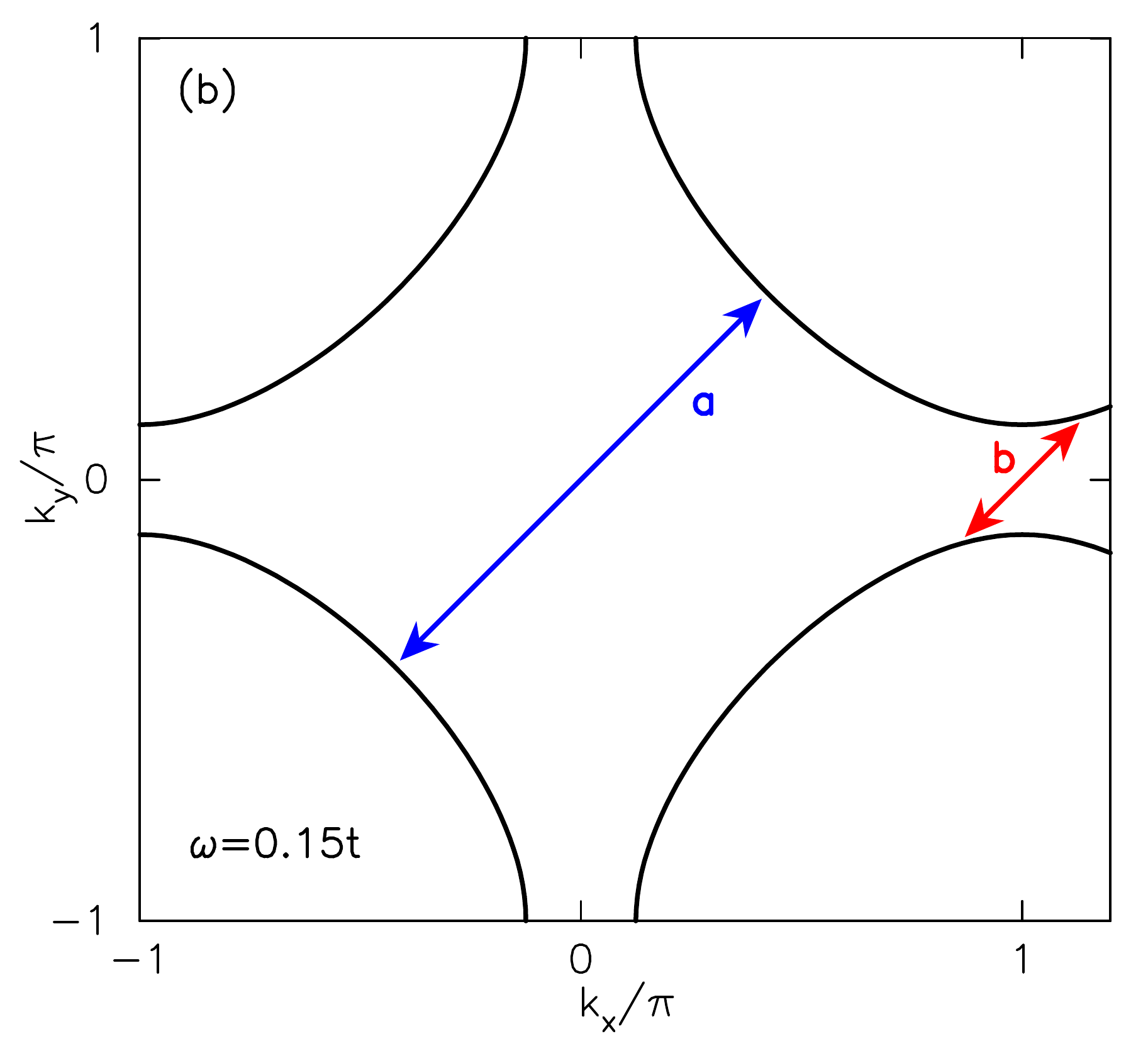}}
\subfigure{
\includegraphics[width=0.43 \columnwidth]{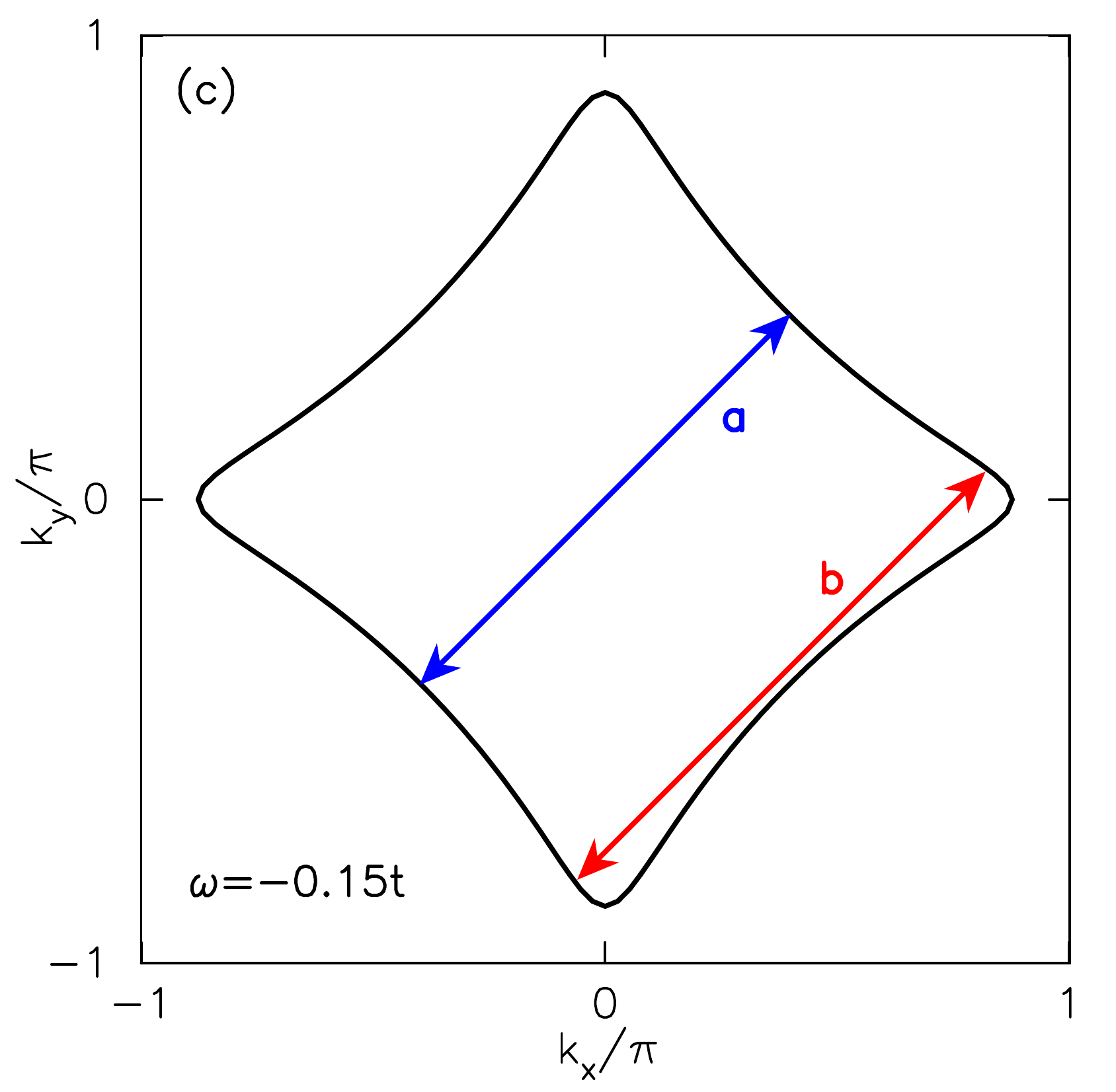}}
\caption{(a) Diagonal and horizontal $q$-cuts for a cuprate like Fermi
surface. In diagonal direction nesting occurs for $q_a$ and $q_b$, while
in horizontal direction nesting occurs for $q_c$.
(b) The $k(\omega)$ surface for $\omega=0.15t$ with diagonal $q$-cuts which
nest for $q_a$ and $q_b$. (c) Similar to (b) for $\omega=-0.15t$.
Here, the nesting vector $q_b$ has changed due to the change
of topology of the $k(\omega)$ surface.\label{fig:2}}
\end{center}
\end{figure}

For weak charge impurity scattering, the wave-vector power spectrum of the local
tunneling density of states depends upon the product of a static impurity
structure factor and the quasi-particle interference response function.
Neglecting vertex corrections, the response function can be written as \cite{Capriotti},
\begin{equation}
  \Lambda(q,\omega)=\int d^2xe^{i{\bf q}\cdot{\bf x}}G(x,\omega)G(-x,\omega)
\label{eq:1}
\end{equation}
with $\omega=eV$. An expression for $\Lambda(q,\omega)$ including vertex corrections
is given in Ref.~\cite{Kivelson}.
Here we consider the QPI response for a given value of
$\omega$ plotted as a function of wave-vector $q$ along certain lines in the
Brillouin zone called $q$-cuts \cite{ref:3}. Examples of such $q$-cuts are
illustrated in Fig.~\ref{fig:1} for a 2D free electron system and in
Fig.~\ref{fig:2}a for a cuprate like band structure. Along a $q$-cut, the QPI
response function Im $\Lambda(q,\omega)$ peaks near wave-vectors $q_{\rm peak}(\omega)$
which connect equi-energy quasi-particle surfaces which have oppositely directed
quasi-particle velocities. As noted, these peaks reflect a dynamic nesting which
depends on the band structure $E(k)$ as well as the quasi-particle self-energy.
For the 2D free electron case the radial $q$-cuts all give the same information,
while for the cuprate-like bandstructure the horizontal $q$-cut (c) shown in
Fig.~\ref{fig:2}a probes the antinodal region of the Fermi surface while the
diagonal $q$-cuts probe both the nodal (a) as well as intermediate regions
(b), which depend on $\omega$, as shown in Fig.~\ref{fig:2}b and c.

As we will discuss, by fitting the $\omega$ dependence of the peak structure in
the QPI, one can extract information on the $\omega$ dependence of the real and
imaginary parts of the self-energy. Both normal and Umklapp peaks associated
with a given $q$-cut provide similar information and in principle the Umklapp
peaks can be used to estimate the $q$ dependent fall-off of the impurity
scattering structure factor.

To begin, we consider the 2D free electron system of Fig.~\ref{fig:1}. In this
case
\begin{equation}
  G(x,\omega)\simeq-i\pi N(0)H^{(1)}_0\left(k(\omega)r\right)
\label{eq:2}
\end{equation}
with $H^{(1)}_0$ the zeroth order Hankel function of the first kind,\cite{Capriotti} $N(0)$ the
single particle density of states, and $k(\omega)$ is determined from the
dispersion relation
\begin{equation}
  \omega=\frac{k^2(\omega)}{2m}-\mu
\label{eq:3}
\end{equation}
Here $\mu=k^2_F/2m$. Carrying out the spatial integration in Eq.~(\ref{eq:1}),
the QPI response function
\begin{equation}
  -\frac{1}{\pi}\ {\rm Im}\,\Lambda(q,\omega)=
	\frac{8\pi N^2(0)}{q}\ {\rm Re}\,\frac{1}{\sqrt{q^2-\left(2k(\omega)\right)^2}}\\
\label{eq:4}
\end{equation}
is found to have a square-root singularity for
\begin{equation}
  q=2k(\omega)=2k\sqrt{1+\frac{\omega}{u}}\simeq2\left(k_F+\frac{\omega}{v_F}\right)
\label{eq:5}
\end{equation}
This wave-vector connects nested equi-energy contours at $k(\omega)$ and
$-k(\omega)$ along the $q$-cut. When impurity scattering is taken into account 
\begin{equation}
  k(\omega)\simeq k_F+\frac{\omega}{v_F}+\frac{i}{2\ell}
\label{eq:6}
\end{equation}
with $\ell$ the mean free path. In this case
the wave vector $k(\omega)=k_1(\omega)+ik_2(\omega)$ contains information on
both the real and imaginary parts of the single particle propagation.

\section{RPA self-energy}

For an interacting system, again neglecting vertex corrections
\begin{equation}
  \Lambda(q,\omega)=\int\frac{d^2k}{(2\pi)^2}G(k,\omega)G(k+q,\omega)
\label{eq:7}
\end{equation}
with
\begin{equation}
  G(k,\omega)=\left[\omega Z(k,\omega) -E(k)\right]^{-1}
\label{eq:8}
\end{equation}
We have set $\Sigma(k,\omega)=\left(1-Z(k,\omega)\right)\omega$ and $E(k)$ is
the bandstructure energy minus the chemical potential $\mu$. For illustration,
we will calculate $Z(k,\omega)=Z_1(k,\omega)+i Z_2(k,\omega)$ for a Hubbard 
model using a random phase approximation (RPA) for 
the spin-fluctuation interaction: \cite{nodalQPL}
\begin{eqnarray}
\label{eq:9}
\lefteqn{  \left(1-Z(k,\omega)\right)\omega =} \\ 
& & -\int\frac{d\omega}{\pi}\int\frac{d^2q}{(2\pi)^2}
	G_0(k+q,\omega+\Omega)
\frac{3}{2}\frac{\bar U^2\chi_0(q,\Omega)}{1-U\chi_0(q,\Omega)}
\nonumber \end{eqnarray}
\begin{equation}
  \chi_0(q,\Omega)=\int\frac{d\omega}{2\pi}\int\frac{d^2k}{(2\pi)^2}
	G_0(k+q,\omega+\Omega)G_0(k,\omega)
\label{eq:10}
\end{equation}
Here, $G_0(k,\omega)=\left[\omega-E(k)+\delta\;{\rm sgn}(\omega)\right]^{-1}$
with a tight-binding bandstructure
\begin{eqnarray}
  E(k) &=& -2t(\cos k_x+\cos k_y)-4t'\cos k_x\cos k_y \nonumber \\
& & -2t''\cos 2 k_x \cos 2 k_y-\mu
\label{eq:11}
\end{eqnarray}
and $t'/t=-0.15$, $t''/t=0.075$, $\mu/t=-0.81$.
The parameters of the bandstructure were taken from tight-binding fits to ARPES data
appropriate for the La$_{2-x}$Sr$_x$CuO$_4$ cuprate near optimum doping $x=0.15$.\cite{Yoshida}
The coupling constants $\bar U/t=3$ and $U/t=1.5$ have been chosen such that
a mass renormalization of $Z_1(\omega=0)\approx 2$ is obtained at the nodal direction
and $Z_1(\omega=0)\approx 3$ at the antinodal direction. For the numerical
calculations a finite broadening of $\delta=0.005 t$ has been used.

\begin{figure}[htbp]
\begin{center}
\subfigure{
\includegraphics[width=0.47 \columnwidth]{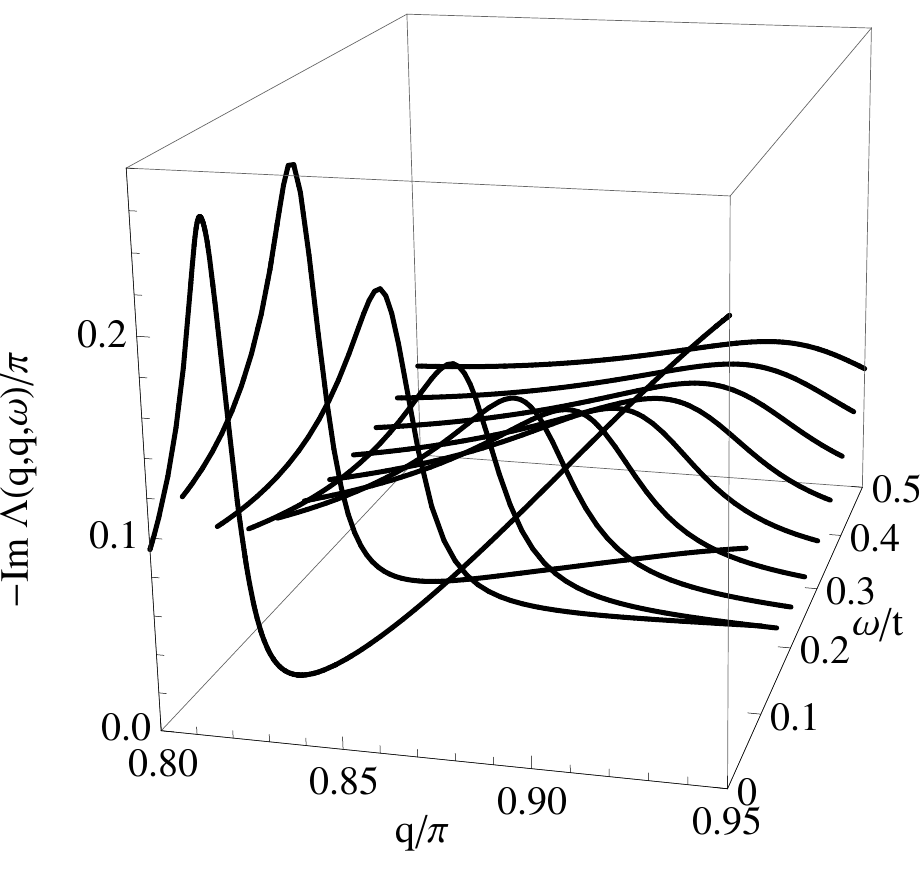}}
\subfigure{
\includegraphics[width=0.47 \columnwidth]{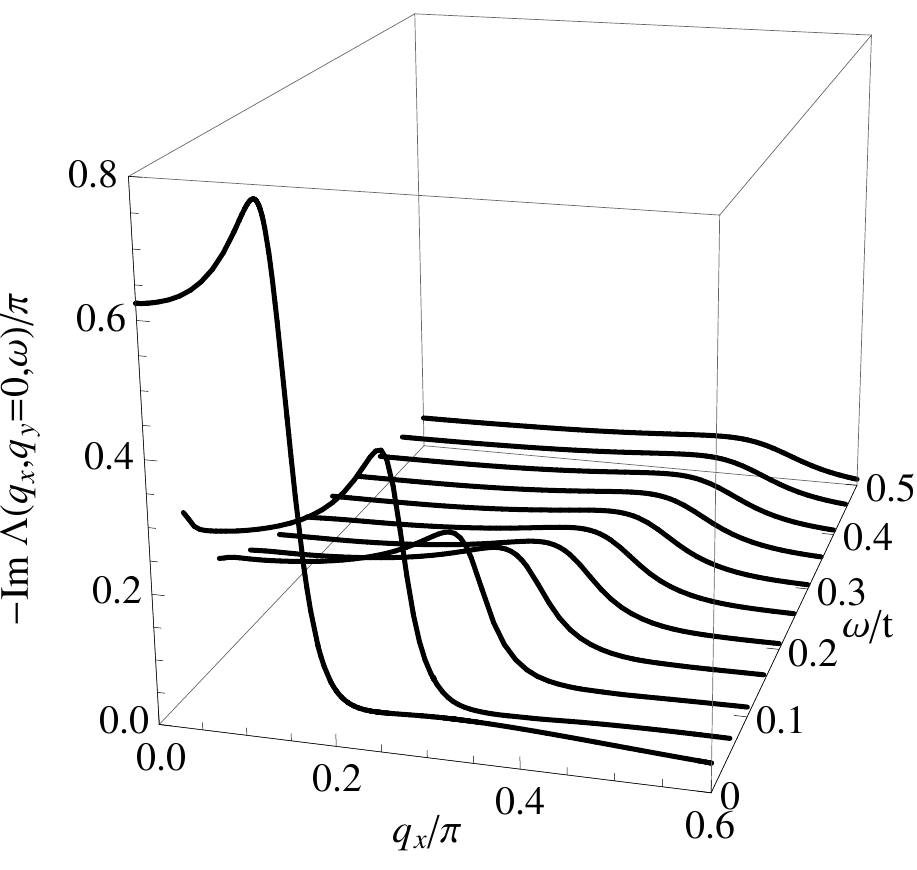}}
\vspace{5mm}
\caption{Plots of $-\frac{1}{\pi}{\rm Im}\,\Lambda(q,\omega)$ versus $q$ for $\omega/t$ values
separated by 0.05 for the diagonal (a) and horizontal (b) $q$-cuts of
Fig.~\protect{\ref{fig:2}}a. As $\omega/t$ increases the QPI peak disperses and
broadens.\label{fig:3}}
\end{center}
\end{figure}

Calculating the RPA self-energy $Z(k,\omega)$ for these parameters and using it
in Eq.~(\ref{eq:7}), we find the QPI response shown in Fig.~\ref{fig:3}.
In the following we will consider $-\frac{1}{\pi}{\rm Im}\,\Lambda(q,\omega)$ as our
``experimental" QPI response and explore how one can extract information about
$Z(k,\omega)$ from it.

\section{Estimation of the self-energy from the
peak position}

From Fig.~\ref{fig:3}, one sees that the response is characterized by a peak
which disperses and broadens as $\omega/t$ increases. The peak in
Fig.~\ref{fig:3}a corresponds to the nesting vector ``a'' in
Fig.~\ref{fig:2}a, while the peak in
Fig.~\ref{fig:3}b corresponds to the nesting vector ``c'' in
Fig.~\ref{fig:2}a. There exists another peak along the diagonal $q$-cut
at smaller values of $q$, which corresponds to the nesting vector ``b''
in Fig.~\ref{fig:2}a, but is not shown here. We do not consider the nesting
vector ``b''
any further, as in the case discussed here it gives similar 
information about the self-energy for antinodal momenta as vector ``c''.

Within a quasi-particle
approximation, one finds that the peaks have similar structure to Eq.~(\ref{eq:4})
\begin{equation}
  -\frac{1}{\pi}\ {\rm Im}\,\Lambda(q,\omega)\sim{\rm Re}\,\frac{1}{\sqrt{2k(\omega)-q}}
\label{eq:12}
\end{equation}
with $k(\omega)=k_1(\omega)+ik_2(\omega)$ determined by
\begin{equation}
  Z_1\left(k_1(\omega),\omega\right)=\frac{E\left(k_1(\omega)\right)}{\omega}
\label{eq:13}
\end{equation}
and
\begin{equation}
  \omega Z_2\left(k_1(\omega),\omega\right)=v_F\left(k_1(\omega)\right)k_2(\omega)
\label{eq:14}
\end{equation}
Here, $v_F\left(k_1(\omega)\right)=\frac{\partial E}{\partial k_\perp}\left(k_1(\omega)\right)$
is the band velocity with the derivative taken perpendicular to the surface where
the $q$-cut crosses the $\omega=E\left( k_1(\omega) \right)$ surface.
The change in sign of the wave vectors in the square-root of Eq.~(\ref{eq:12})
relative to Eq.~(\ref{eq:4}) arises from the change in sign of the Fermi surface
curvature.

\begin{figure}[htbp]
\begin{center}
\subfigure{
\includegraphics[width=0.45 \columnwidth]{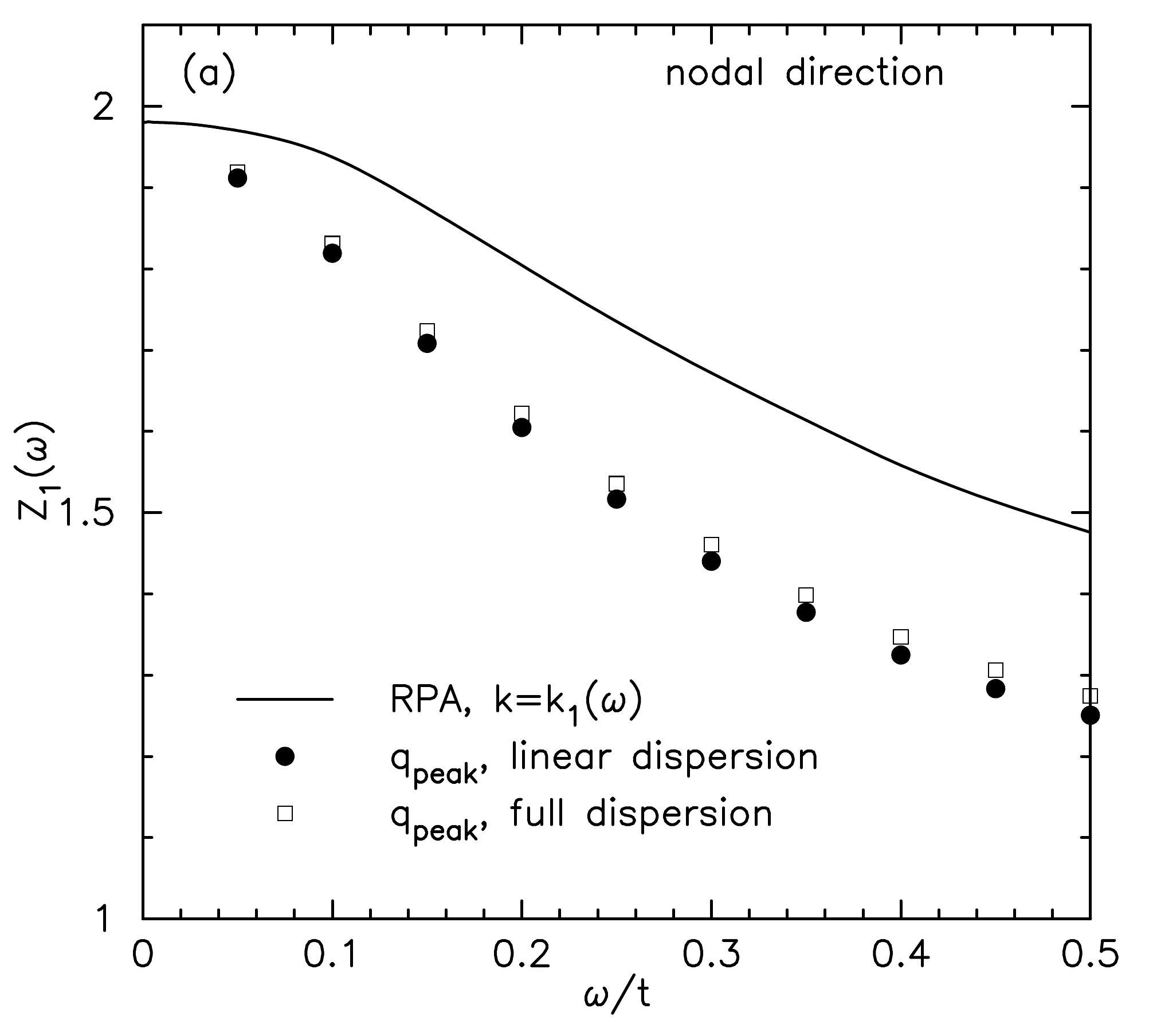}}
\subfigure{
\includegraphics[width=0.45 \columnwidth]{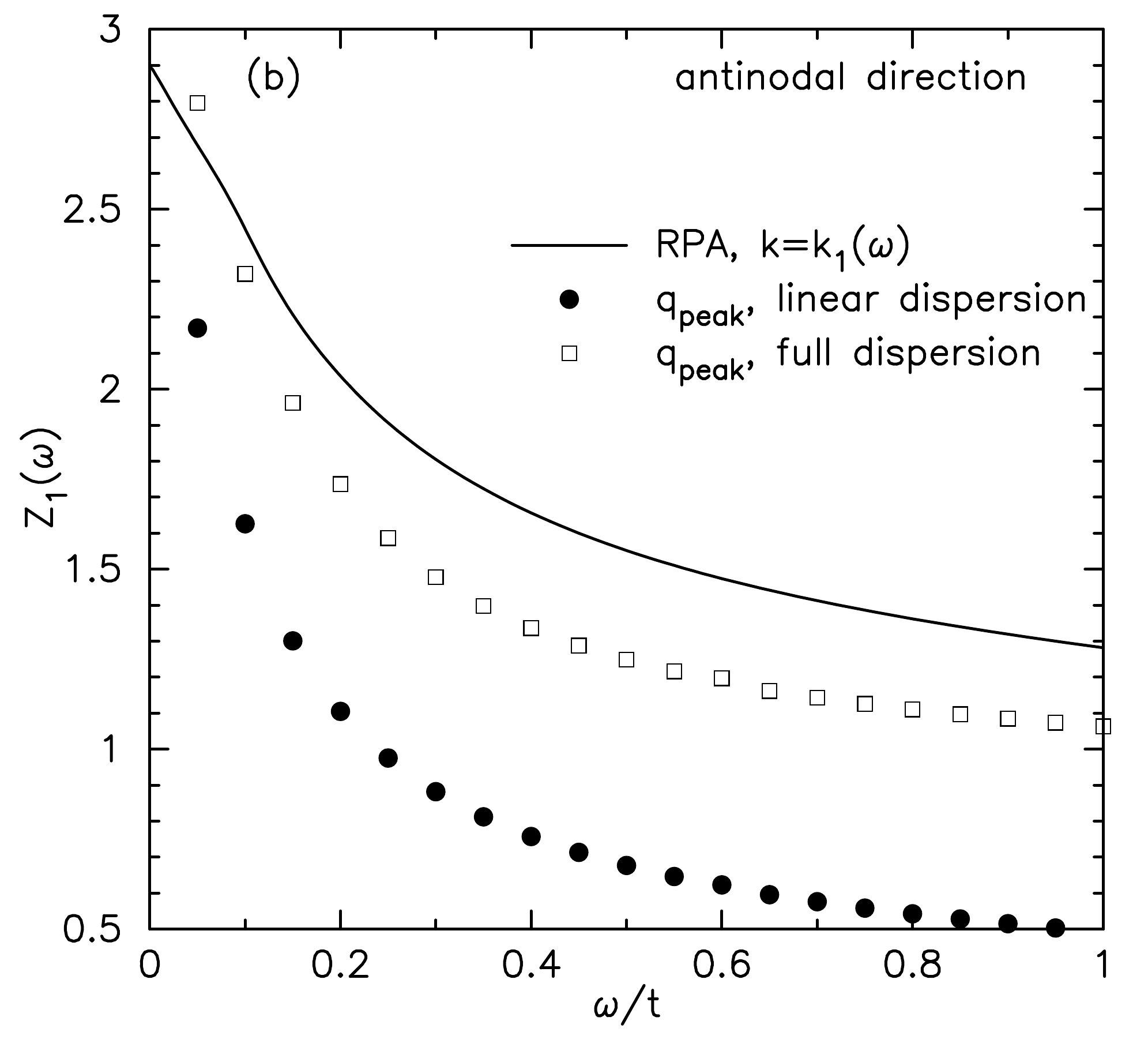}}
\caption{Comparison of $E\left(q_{\rm peak}(\omega)/2\right)/\omega$
(open squares) and its linear form $v_F\left(q_{\rm peak}(\omega)/2-k_F\right)/\omega$
(solid circles) with $Z_1(k_1(\omega),\omega)$ (solid curve) for (a) nodal and (b) antinodal
$q$-cuts.
\label{fig:4}}
\end{center}
\end{figure}

If one takes the peak value $q_{\rm peak}(\omega)$ as an estimate of $2k_1(\omega)$,
neglects the imaginary part and uses a linearized dispersion
$E\left(k_1(\omega)\right) \approx v_F \left( k_1(\omega) - k_F \right)$
one finds approximately
\begin{equation}
  Z_1(k_F,\omega)\simeq\frac{v_F}{\omega}\left(\frac{q_{\rm peak}(\omega)}{2}-k_F\right).
\label{eq:15}
\end{equation}
If $q_{\rm peak}(\omega)$ exceeds the region over which a linear approximation of the
dispersion is appropriate, then one needs to use the full dispersion and
\begin{equation}
  Z_1(k_F,\omega)\simeq\frac{E\left(q_{\rm peak}(\omega)/2\right)}{\omega}.
\label{eq:16}
\end{equation}
In Fig.~\ref{fig:4}
the results for $Z_1(\omega)$ obtained using $q_{\rm peak}(\omega)/2$ as an
estimate of $k_1(\omega)$ for both the diagonal and horizontal $q$-cuts are compared
with $Z_1\left(k_1(\omega),\omega\right)$ (solid curves) obtained from Eq.~(\ref{eq:9}).
For the nodal direction, $E(k)$ is well approximated by its linearized $v_F(k-k_F)$
form while for the antinodal direction a linear approximation fails due to the closeness
of the Fermi level to the saddle point of the band at $(0,\pi)$. In
this case it is necessary
to use the full band dispersion $E(k)$. In both cases, using $q_{\rm peak}(\omega)$
underestimates $2k_1(\omega)$ and the resulting $Z_1(\omega)$ falls below the
self-energy used in the calculation of $\Lambda(q,\omega)$.

\section{Fitting of the QPI peaks}

The problem with using the peak of the QPI $q$-cut to estimate $Z_1(\omega)$ is
that $k(\omega)$ in Eq.~(\ref{eq:12}) has both real and imaginary parts. Thus a
better alternative is to fit the QPI peak to the square-root form of Eq.~(\ref{eq:12})
and extract a $k_1(\omega)$ and $k_2(\omega)$ for each $\omega$ $q$-cut.
\begin{figure}[htbp]
\begin{center}
\includegraphics[width=0.45 \columnwidth]{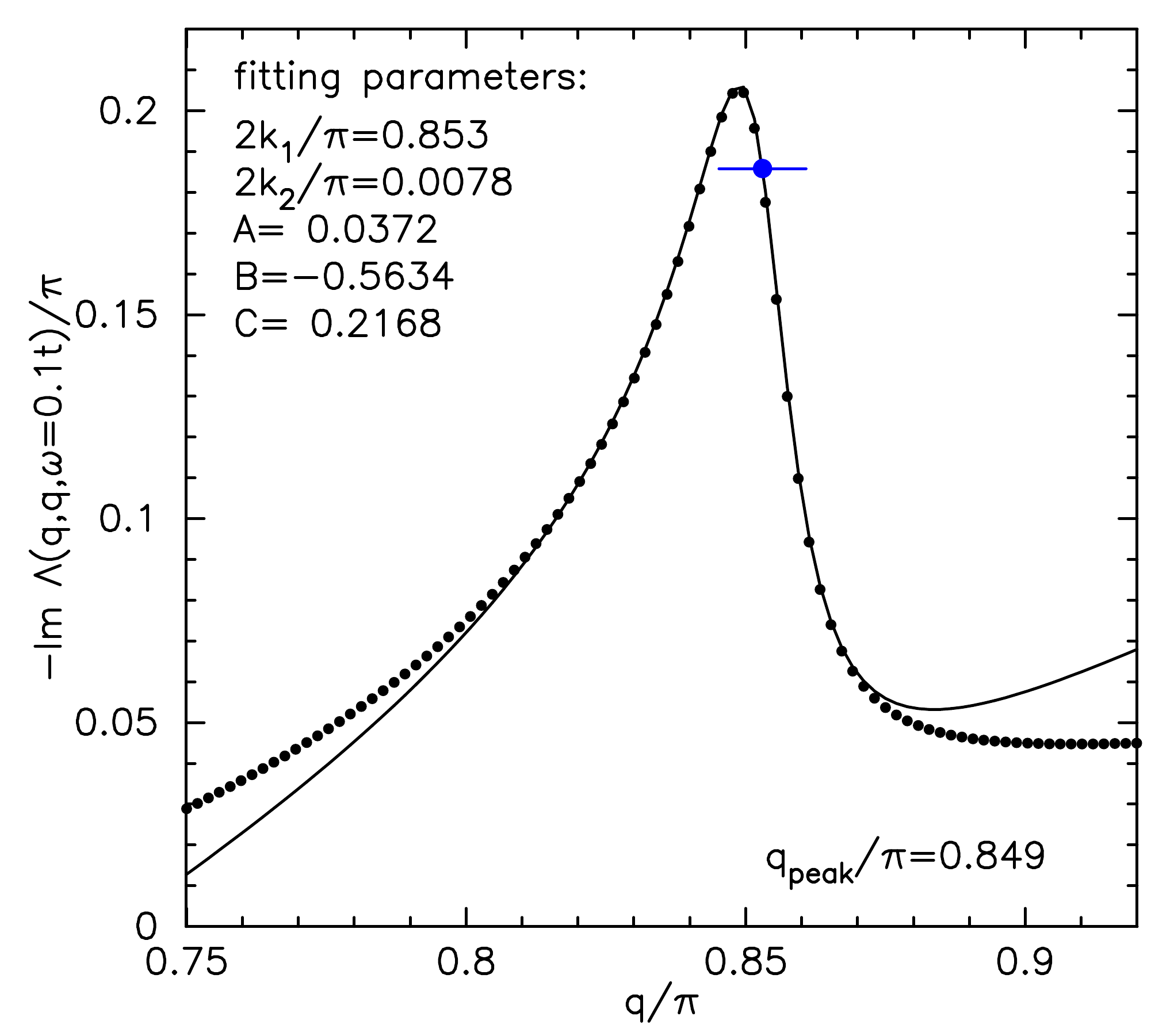}
\caption{Fitting $-\frac{1}{\pi}{\rm Im}\,\Lambda(q,\omega=0.1t)$ along a diagonal $q$-cut
using Eq.~(\protect{\ref{eq:17}}) to obtain $k_1(\omega)$ and $k_2(\omega)$.
\label{fig:5}}
\end{center}
\end{figure}
Figure~\ref{fig:5} shows the results of a fit to a form
\begin{equation}
  -\frac{1}{\pi}\ {\rm Im}\,\Lambda(q,\omega)={\rm
   Re}\,\frac{A}{\sqrt{2k_1(\omega)+i 2k_2(\omega)-q}}+B+Cq \, .
\label{eq:17}
\end{equation}
Here, $\omega=0.1t$ along a diagonal $q$-cut is shown.
A finite $q$ range of $[ 0.825, 0.875 ] \pi$ has been used for the fitting.
As can be seen in Fig.~\ref{fig:3} the peaks for a cuprate-like
bandstructure appear to have a linear ``background'' behind the
smeared square-root singularity. For that reason we found it necessary to
include a linear background $B+Cq$ in the fitting formula Eq.~(\ref{eq:17}), 
which improves the determination of $k_1$ and $k_2$. The blue dot in 
Fig.~\ref{fig:5} denotes the position of the extracted $2k_1$ and the blue bar
the width $\pm 2k_2$ for this particular energy $\omega$. As this plot shows, the
actual position of $2k_1$ is slightly off from the peak position $q_{\rm peak}$
to the right. The reason for this is the asymmetric line shape of the
square-root singular QPI response function Eq.~(\ref{eq:12}).
As seen below, the fitted values of $k_1$ and $k_2$ allow a much more precise 
extraction of $Z_1$ and $Z_2$ than the peak position $q_{\rm peak}/2$.

To get precise values of $k_1$ and $k_2$ from the fit we found it necessary
to restrict the fitting to a finite $q$ range around the peak position
to avoid the fit being spoiled by values away from the peak, where
the fitting formula Eq.~(\ref{eq:17}) is not valid anymore.
To get a good coverage of the peak we have chosen the following
$q$ ranges: $[ q_{\rm peak}-0.02\pi(1+|\omega|/t), q_{\rm peak}+0.02\pi(1+2|\omega|/t) ]$
in the nodal direction and $[ q_{\rm peak}-0.02\pi(1+4|\omega|/t), 
q_{\rm peak}+0.02\pi(1+8|\omega|/t) ]$
in the antinodal direction. These $q$ ranges account for a minimum range of
$\Delta q = 0.04 \pi$, they increase with increasing $|\omega|$ to account for
the fact that the peaks are getting broader at higher frequencies, and the
larger range in the antinodal direction accounts for the smaller Fermi velocity
in this direction which leads to larger peak widths. Also note that we have
chosen the $q$ range asymmetrically around the peak position, as $2k_1$ is
always larger than $q_{\rm peak}$.

After $k_1(\omega)$ and $k_2(\omega)$ have been extracted from these fits,
for the linear dispersion approximation we will compare
\begin{equation}
  v_F\left(k_1(\omega)-k_F\right)/\omega
\label{eq:18}
\end{equation}
and
\begin{equation}
  v_Fk_2(\omega)
\label{eq:19}
\end{equation}
with $Z_1\left(k_1(\omega),\omega \right)$ and $\omega Z_2\left(k_1(\omega),\omega\right)$,
respectively. In the following all of our estimates will be compared with
$Z\left(k_1(\omega),\omega\right)$ since the self-energy does have a weak
$k$ dependence. If the dynamic range is such that the non-linearity of
the dispersion is important, then the comparison will be with
\begin{equation}
  Z_1\left(k_1(\omega),\omega\right)=E\left(k_1(\omega)\right)/\omega
\label{eq:20}
\end{equation}
and
\begin{equation}
  \omega Z_2\left(k_1(\omega),\omega\right)=\left(\frac{\partial E}{\partial k_\perp}\left(k_1(\omega)\right)\right)k_2(\omega)
\label{eq:21}
\end{equation}
In the following plots, the solid curves are the RPA self energy evaluated at
$k=k_1(\omega)$ with $k_1(\omega)$ in this case obtained from the self-energy
calculation Eq.~(\ref{eq:9}). Our goal is to see how well one can extract the
solid curves from the QPI response $- \frac{1}{\pi}\ {\rm Im}\,\Lambda(q,\omega)$.

\begin{figure}[htbp]
\begin{center}
\subfigure{
\includegraphics[width=0.45 \columnwidth]{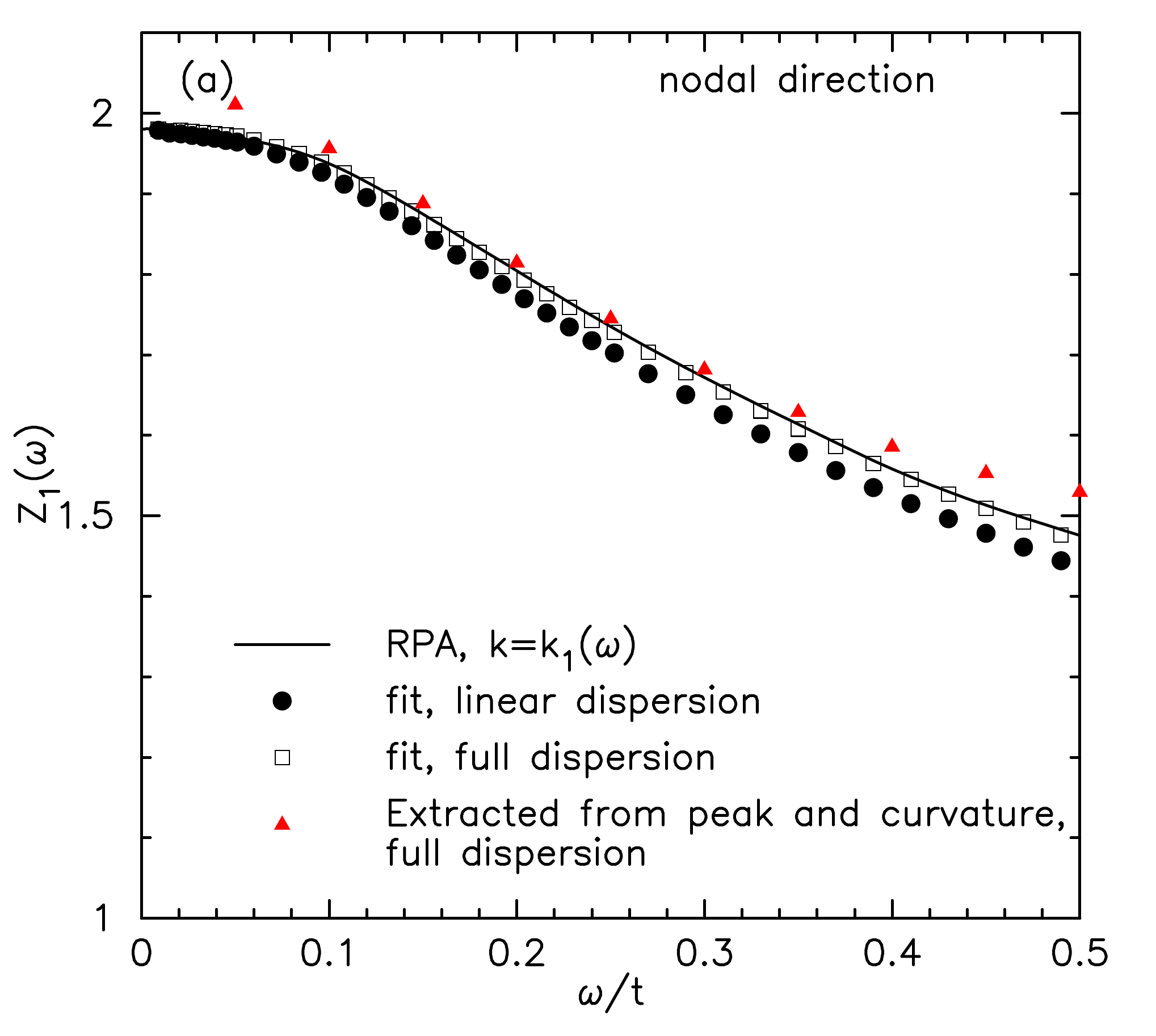}}
\subfigure{
\includegraphics[width=0.45 \columnwidth]{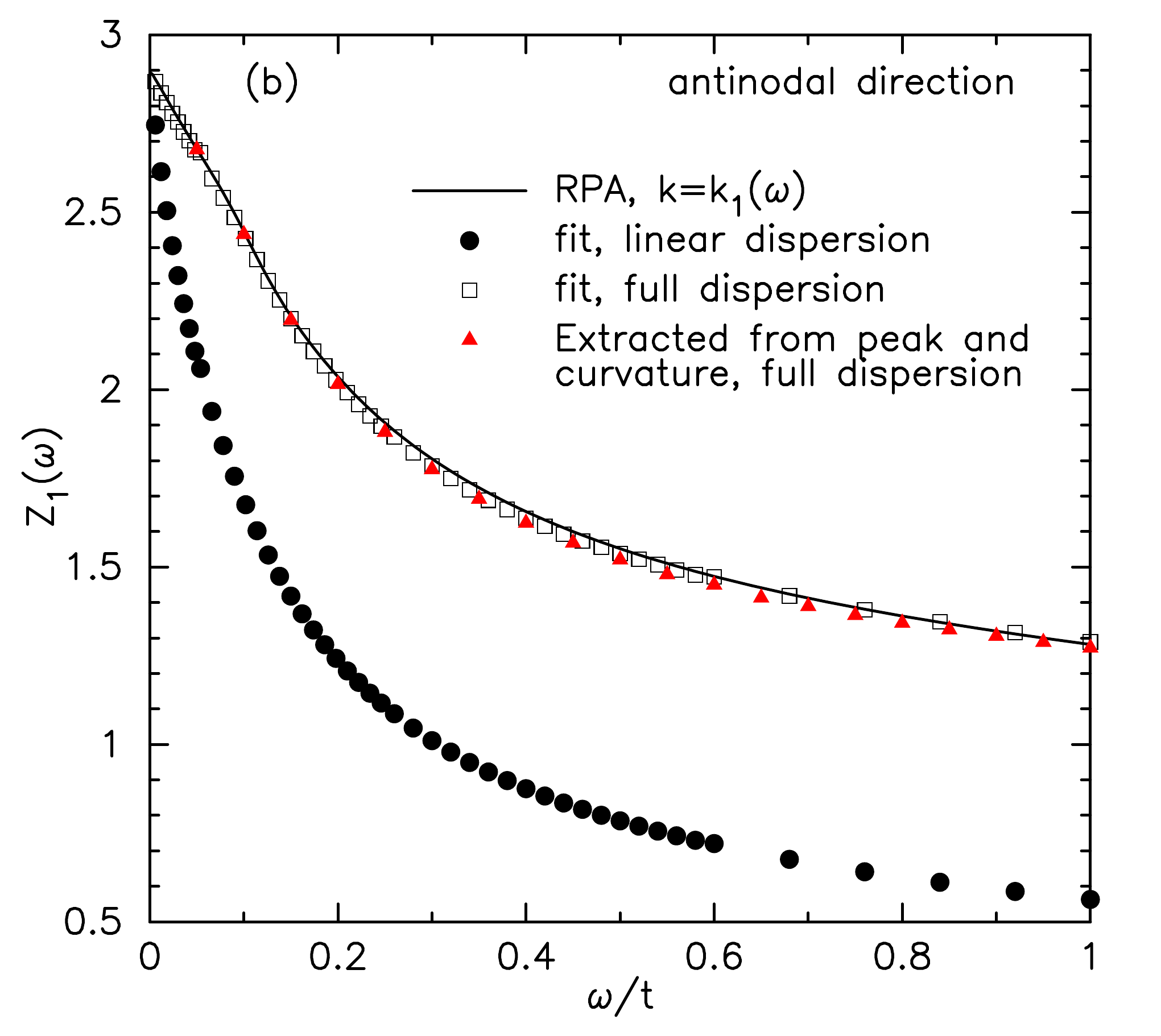}}
\end{center}
\caption{Comparisons of $E\left(k_1(\omega)\right)/\omega$ (open squares) and
its linearized form $v_F\left(k_1(\omega)-k_F\right)/\omega$ (solid circles)
with $Z_1\left(k_1(\omega),\omega\right)$ (solid curve) for (a) nodal and (b)
antinodal $q$-cuts. Here $k_1(\omega)$ used in $E(k_1(\omega))$ and $v_F(k_1(\omega)-k_F)$
is obtained from fitting Eq.~(\protect{\ref{eq:17}}). The (red) triangles are obtained
when $k_1(\omega)$ is extracted using the peak and normalized curvature,
Eqs.~(\protect{\ref{eq:22}}) and (\protect{\ref{eq:23}}).\label{fig:6}}
\end{figure}

\begin{figure}[htbp]
\begin{center}
\subfigure{
\includegraphics[width=0.45 \columnwidth]{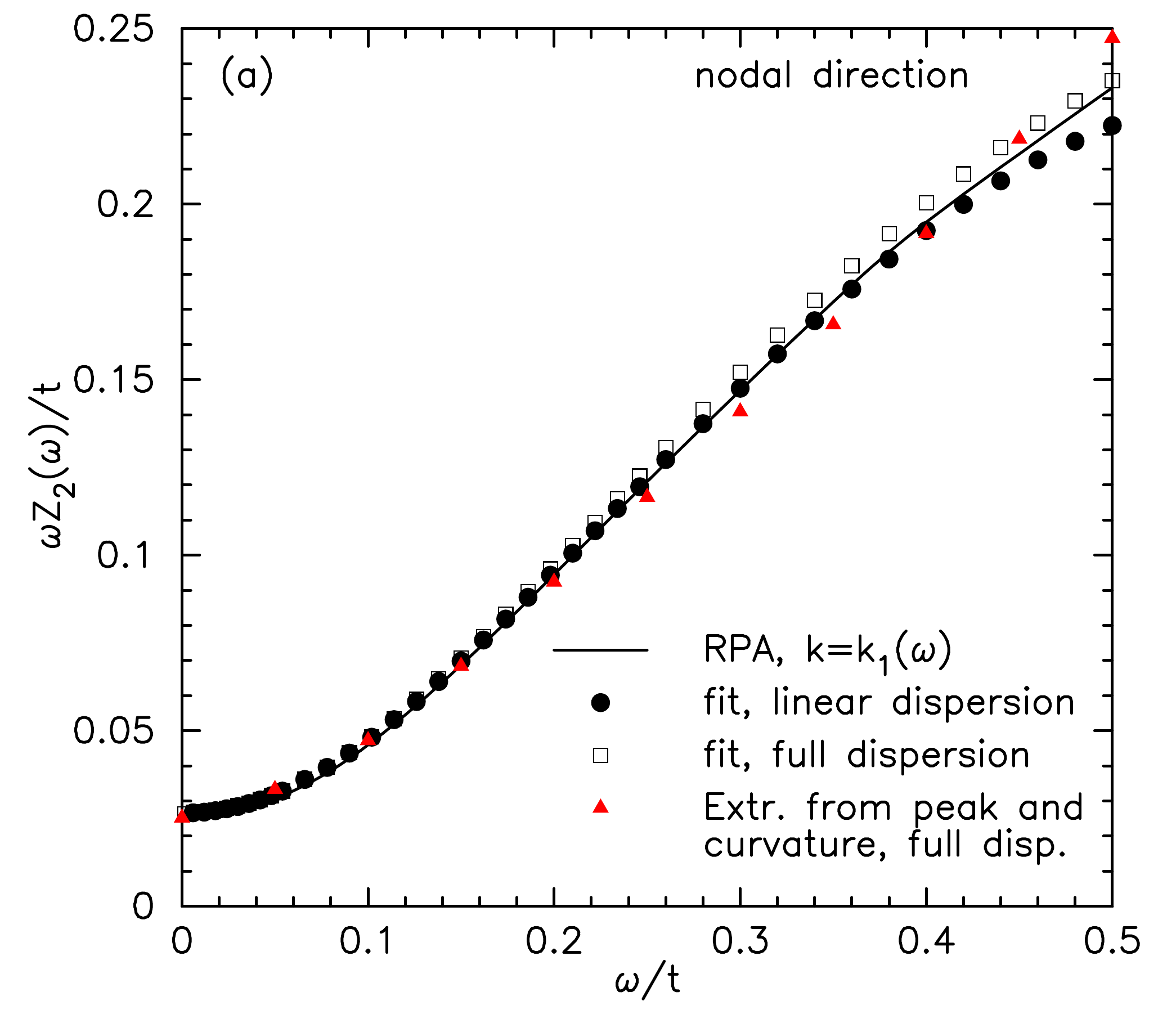}}
\subfigure{
\includegraphics[width=0.45 \columnwidth]{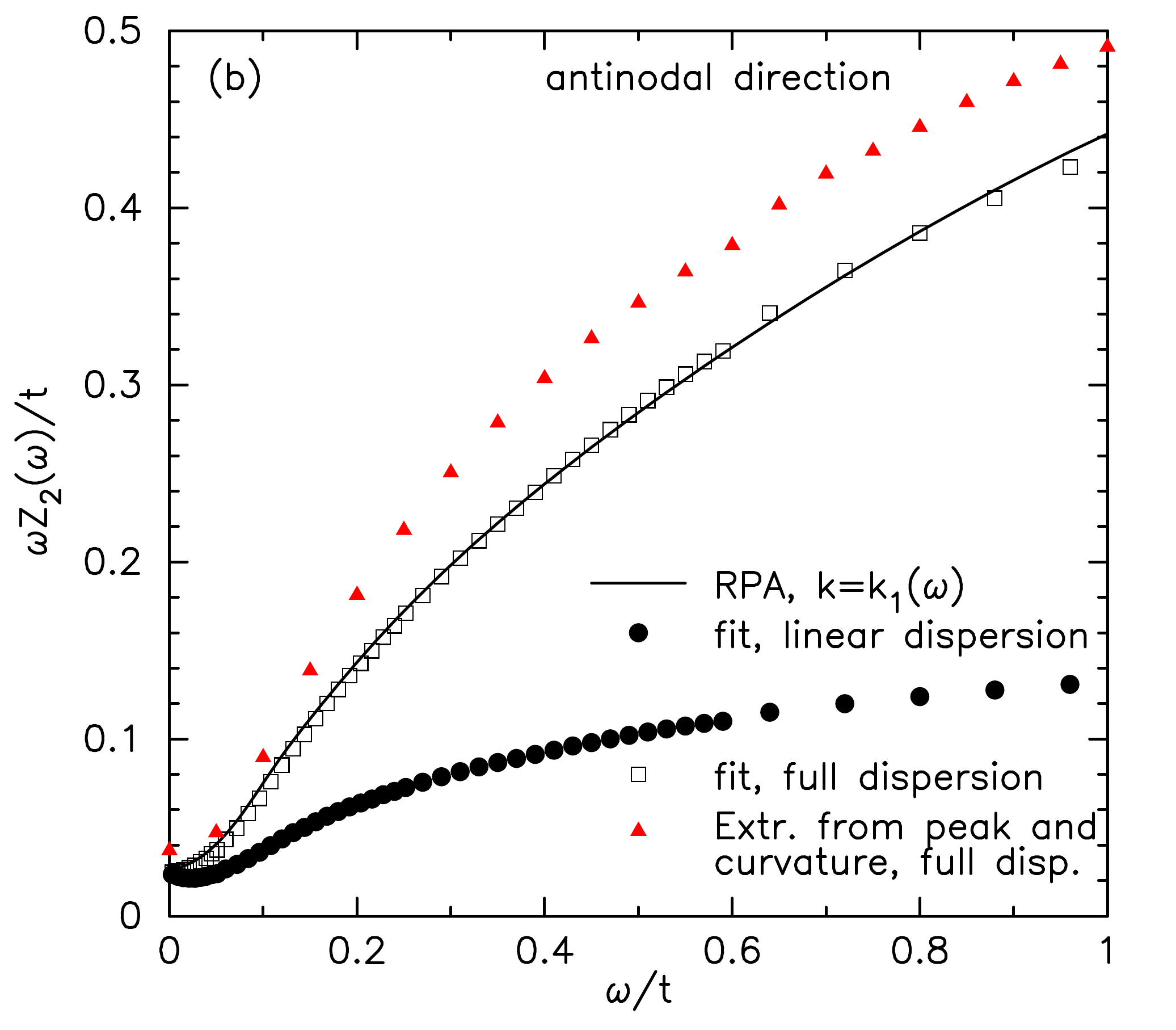}}
\end{center}
\caption{Comparisons of $\frac{\partial E}{\partial k_\perp}\left(k_1(\omega)\right)k_2(\omega)$
(open squares) and its linearized form $v_Fk_2(\omega)$ (solid circles) with
$\omega Z_2\left(k_1(\omega),\omega\right)$ (solid curve) for (a) nodal and (b)
antinodal $q$-cuts. Here $k_1(\omega)$ and $k_2(\omega)$ were obtained by fitting
Eq.~(\protect{\ref{eq:17}}). The (red) triangles were again obtained using
$k_1(\omega)$ and $k_2(\omega)$ extracted using the peak and normalized curvature.
\label{fig:7}}
\end{figure}

As shown in Fig.~\ref{fig:6}a and  Fig.~\ref{fig:7}a for the nodal case, 
useful estimates of
$Z_1\left(k_1(\omega),\omega\right)$ and $Z_2\left(k_1(\omega),\omega\right)$
can be obtained using $\left(k_1(\omega),k_2(\omega)\right)$ and a linear
approximation of the dispersion.
Similarly in Fig.~\ref{fig:6}b and  Fig.~\ref{fig:7}b one sees 
that if $\left(k_1(\omega),k_2(\omega)\right)$
can be extracted by fitting ${\rm Im}\,\Lambda(q,\omega)$ for the
antinodal case one can
again obtain useful estimates of the self-energy. However, in the antinodal
case it is important to use the full bandstructure $E(k)$.

\section{Extraction using peak position and normalized
curvature}

While fitting ${\rm Im}\,\Lambda(q,\omega)$ with Eq.~(\ref{eq:17}) provides a way
of extracting $k_1(\omega)$ and $k_2(\omega)$, as well as supports the validity
of the approximation Eq.~(\ref{eq:12}) near the peak position, one would like to 
have a more direct procedure,
which avoids choosing a finite $q$ range for the fit.
From Eq.~(\ref{eq:12}) one finds that the peak occurs for
\begin{equation}
  q_{\rm peak}=2k_1(\omega)+2k_2(\omega)/\sqrt{3}
	\label{eq:22}
\end{equation}
and at the peak, the normalized curvature
\begin{equation}
  \frac{\left.\frac{d^2\ {\rm Im}\,\Lambda(q,\omega)}{d q^2}\right|_{q_{\rm peak}}}{{\rm Im}\,\Lambda(q_{\rm peak},\omega)}
	=-\frac{9}{16}\frac{1}{(2k_2)^2}
  \label{eq:23}
\end{equation}

Results obtained for $Z_1$ and $Z_2$ using Eqs.~(\ref{eq:22}) and (\ref{eq:23})
are shown as the red triangles in Figs.~\ref{fig:6} and \ref{fig:7}.
While this way of extracting $k_1(\omega)$ and $k_2(\omega)$ from the QPI response
is less accurate than fitting Eq.~(\ref{eq:17}), it can provide reasonable
results. The $B+Cq$ background must be removed from ${\rm Im}\,\Lambda(q,\omega)$
in estimating the normalized curvature, Eq.~(\ref{eq:23}).

\section{Conclusions}

While we have been able to extract from ${\rm Im}\,\Lambda(q,\omega)$ the
self-energy that went into the Green's functions used to calculate it, these
results illustrate some of the challenges and limitations one faces. Extracting
$k_1(\omega)$ and $k_2(\omega)$ will clearly become more difficult as $\omega$
increases and the peak broadens and decreases in amplitude \cite{ref:7}. As seen
for the diagonal $k_x=k_y$ cut in Fig.~\ref{fig:2}, there can also be multiple
peaks associated with a given $q$-cut. For $\omega>0$, these peaks are well
separated in $q$. However, as seen in Fig.~\ref{fig:2}c, for this Fermi surface
there are problems for $\omega<0$, where the $q_a$ and $q_b$ peaks approach
each other for $\omega<0$. In addition to $k_1(\omega)$ and $k_2(\omega)$ we
used information on the bare bandstructure $E(k)$ which is not a measured
quantity. For the diagonal $q$-cut it appears that the band Fermi velocity $v_F$
would be sufficient, and in principle one might hope that at large values of
$\omega$ one could estimate the bare $v_F$. However, by these energies one is
typically out of the linear region of dispersion. Thus one needs to make a
reasonable estimate for $E(k)$ based on band theory.

Finally, there is the $q$-dependence of the impurity scattering form factor and
the effect of vertex corrections. \cite{Kivelson} The scattering form factor will reduce the
amplitude of the QPI response as $q$ increases but should not lead to significant
shifts of the $k_1(\omega)$ and $k_2(\omega)$ values provided its characteristic
momentum is large compared with $k_2(\omega)$. While the vertex $\Gamma(k,q)$
can introduce additional structure, the continuity of the peak associated with
the interference between the two propagators as $\omega$ increases along with
the short range nature of the vertex corrections should generally limit its
effect on $k_1(\omega)$ and $k_2(\omega)$.

\section*{Acknowledgments}

DJS acknowledges the support of the Center for Nanophase Materials Science at
ORNL, which is sponsored by the Division of Scientific User Facilities, U.S. DOE.



\begin{thebibliography}{99}
\bibitem{ref:7} By going into the superconducting state one can obtain a sharper response.
In this case one will need to take into account the superconducting gap. However
at energies large compared with $|\Delta(k)|$ the corrections to $Z$ will be of
order $|\Delta/\omega|^2$.
\end{thebibliography}
\end{document}